\documentclass[twoside]{dis09}
\usepackage[latin1]{inputenc}
\usepackage[dvips]{graphicx,epsfig,color}
\usepackage{wrapfig,rotating}
\usepackage{amssymb,amsmath,array}

\begin{document}
\newcommand{\pom}{{I\!\!P}}
\newcommand{\reg}{{I\!\!R}}
\newcommand{\tmod}{\ensuremath{|t|}}
\newcommand{\qsq}{\ensuremath{Q^2} }
\newcommand{\gevsq}{\ensuremath{~\mathrm{GeV}^2} }
\newcommand{\ptgamma}{\ensuremath{P_{T}^{\gamma}} }
\newcommand{\alpeff}{\alpha_s^{BFKL}}
\newcommand{\wfitres}{\ensuremath{\delta=2.73 \pm 1.02~\mathrm{(stat.)}
^{+0.56}_{-0.78}~\mathrm{(syst.)}}}
\newcommand{\tfitres}{\ensuremath{n=2.60 \pm 0.19~\mathrm{(stat.)}
^{+0.03}_{-0.08}~\mathrm{(syst.)}}}
\newcommand{\alphasfitres}{\ensuremath{\alpha_S}^{Fit}=0.26 \pm 0.10
~\mathrm{(stat.)} ^{+0.05}_{-0.07}~\mathrm{(syst.)}}
\newcommand{\xpom}{\ensuremath{x_{\pom}}}
\newcommand{\ypom}{\ensuremath{y_{\pom}}}

\title{Measurement of Diffractive Scattering of Photons with Large Momentum Transfer at HERA}

\author{Tom\'a\v{s} Hreus (On behalf of the H1 Collaboration)
%
%
\vspace{.3cm}\\
%
Universit\'e Libre de Bruxelles - I.I.H.E. \\
Boulevard du Triomphe, CP230, 1050 Bruxelles, Belgium
%
}

\maketitle

\begin{abstract}
The first measurement of diffractive scattering of quasi-real
photons with large momentum transfer $\gamma p \rightarrow \gamma Y$,
where $Y$ is the proton dissociative system, is made using the H1
detector at HERA.
Single differential cross sections are measured as a function of $W$, the
incident photon-proton centre of mass energy,
and $t$, the square of the four-momentum
transferred at the proton vertex.
The $W$ dependence is well described by a 
perturbative QCD model using a leading logarithmic approximation
of the BFKL evolution, whereas 
the measured $\tmod$ dependence is harder than predicted.
\end{abstract}

\section{Introduction}
\begin{wrapfigure}{r}{0.5\columnwidth}
\centerline{
\begin{picture}(180,120)
  \put(0,-10){\includegraphics[width=0.45\columnwidth,clip=]{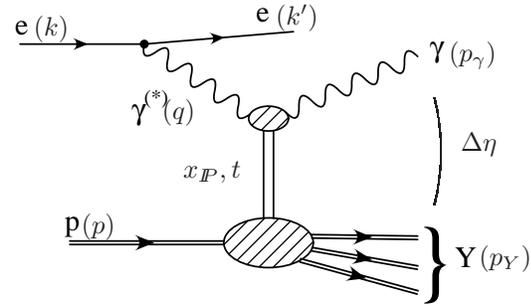}}
  \put(65,50){$\xpom, t$}
  \put(8.5,104){$(k)$}
  \put(100,108){$(k')$}
  \put(165.5,94){$(p_{\gamma})$}
  \put(57,72){$(q)$}
  \put(26.5,29){$(p)$}
  \put(177,17){$(p_Y)$}
  \put(170.0,60.0){$\Delta \eta$}
  \put(160,60){\qbezier(0,20.0)(5,0)(0,-20.0)}
 \end{picture}
}
\caption{Schematic illustration of the $e p \rightarrow e \gamma Y$ process.}
\label{Fig:PhotonDiag1}
\end{wrapfigure}
The study at the $ep$ collider HERA of exclusive diffractive processes
in the presence of
a hard scale provides insight into the parton dynamics of
the diffractive exchange. 
The four-momentum squared transferred at the
proton vertex, $t$, provides a relevant scale to
investigate the application of perturbative Quantum Chromodynamics
(pQCD) for $\tmod \gg \Lambda^2_{\rm QCD}$ \cite{Forshaw:1997wn}.
The diffractive photon-proton scattering, $\gamma p \to \gamma Y$
at large $|t|$
gives an experimentally clean and almost fully perturbatively calculable 
process to study the parton dynamics.
Its cross section 
measurement \cite{:2008cq} is performed at HERA by studying the reaction
$e^+  p \to e^+  \gamma Y$ in the photoproduction regime 
(initial photon virtualities $\qsq < 0.01$~GeV$^2$) and at $4<\tmod<36$~GeV$^2$, 
with a large
rapidity gap between the final state photon and the proton dissociative
system~$Y$ (as illustrated in Fig.~\ref{Fig:PhotonDiag1})
and the $\gamma p$ centre of mass energy in the range 
$175<W<247~\mathrm{GeV}$.

\begin{wrapfigure}{r}{0.5\columnwidth}
\centerline{\includegraphics[width=0.45\columnwidth,clip=]{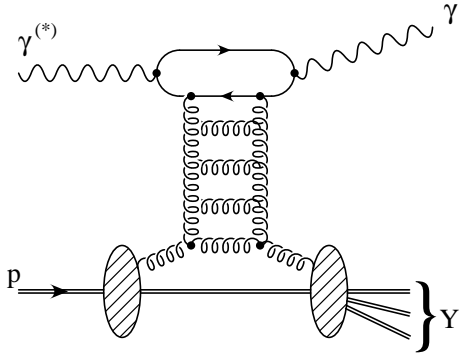}}
\caption{Illustration of the $\gamma^{(*)} p \rightarrow \gamma Y$ process in a
  LLA BFKL approach.}
\label{Fig:PhotonDiag2}
\end{wrapfigure}

Diffractive photon scattering can be modelled in the proton rest
frame by the fluctuation of the incoming photon into a $q \bar q$ pair,
which is then involved in a hard interaction with the
proton via the exchange of a colour singlet state.
In the leading logarithmic approximation (LLA), the colour
singlet exchange is
modelled by the effective exchange of a gluon ladder (Fig.~\ref{Fig:PhotonDiag2}),
described at sufficiently low values of Bjorken $x$ by
the BFKL \cite{Kuraev:1977fs} approach.
In the LLA BFKL model the gluon ladder couples to a single parton
within the proton.
Due to the quasi-real nature of the incoming photon, 
the transverse momentum of the final state photon,
$\ptgamma$, is entirely transferred by the gluon ladder to the parton
in the proton.
The separation in rapidity between the parton scattered by the gluon
ladder and the final state
photon is given by $\Delta \eta \simeq {\rm log} (\hat s / (\ptgamma)^2)$,
where $\hat s$ is the invariant mass of the system formed by the
incoming photon and the struck parton.

In addition to the usual DIS kinematic variables,
the longitudinal momentum fraction of the diffractive exchange with respect to the proton, $\xpom$ and
the elasticity of the $\gamma p$ interaction, $\ypom$, are defined as 
(Following the notation in Fig.~\ref{Fig:PhotonDiag1})
\begin{equation}
 \nonumber
\xpom = \frac{q \cdot (p-p_Y)}{q \cdot p} \, ,~~~~~~
\ypom = \frac{p \cdot (q-p_{\gamma})}{p \cdot q}.
\end{equation}

\section{Experimental procedure}

The data, corresponding to an integrated luminosity of $46.2$ pb$^{-1}$,
were collected with the H1 detector during
the $1999-2000$ running period, when positrons of energy
$E_e=27.6~\rm{GeV}$
collided with protons of energy $E_p=920~\rm{GeV}$. 
The detailed description of the H1 detector can be found in \cite{Abt:1996hi}.
The quantity $W$ is calculated from
the measured energy of the scattered positron, $E_{e'}$, using the
relation $W^2 \simeq ys \simeq s(1-E_{e'}/E_e)$
with the relative resolution $4\%$,
where $s=(k+p)^2$ is the $ep$ centre of mass energy squared.
The scattered positron is detected in a small calorimeter 33 m downstream
the main detector.
The photon candidate is detected as a cluster in the electromagnetic section 
of the calorimeter, with no associated track
and is furthermore required
to have an energy $E_{\gamma}>8~\mathrm{GeV}$ and
a transverse momentum $\ptgamma>2~\mathrm{GeV}$.
The kinematic variable $|t|$ is reconstructed as $|t| \simeq (\ptgamma)^2$, 
with the resolution of 11\%.
The hadronic final state $Y$ is reconstructed using a combination of
tracking and calorimetric information.
The event inelasticity of the $\gamma p$ interaction is reconstructed as
$\ypom \simeq \sum_Y(E-P_z)/[2(E_e - E_{e'})]$, where
the summation is performed on all
detected hadronic final state particles in the event.
Diffractive events are selected by requiring $\ypom < 0.05$,
ensuring a large pseudorapidity gap, $\Delta \eta$, between the
photon and the proton dissociative system~$Y$, since
$\ypom \simeq e^{-\Delta \eta}$.

After applying all selection criteria, described in a more detail in 
\cite{:2008cq,Hreus:2008zz}, 240 events remain in the data sample.

\section{Models}
The {\sc HERWIG} 6.4 MC event generator is used to
simulate the diffractive high $\tmod$ photon scattering using the LLA BFKL
model \cite{Ivanov:1998jw,Evanson:1999zb,Cox:1999kv}.
At leading logarithmic accuracy there are two independent free
parameters in the BFKL calculation: the value of the strong coupling
$\alpha_s$ and the scale, $c$,
which defines the leading logarithms in the
expansion of the BFKL amplitude, $\ln(W^2/(c \ \tmod))$.
In exclusive production of vector mesons, the scale parameter $c$ is
related to the vector meson mass.
In the case of diffractive photon scattering, the unknown scale
results in the absence of a normalisation prediction of the
cross section. In the calculations considered here,
the running of $\alpha_s$ as a function of the scale is ignored and
will henceforth be referred to as $\alpeff$.
The choice of $\alpeff=0.17$ is used for the simulation for this measurement.

In the asymptotic approximation of the calculations
\cite{Cox:1999kv}, the W dependence of the cross section follows a power-law,
$\sigma(W) \sim  W^{4\omega_0}$,
where the exponent is given by the choice of $\alpeff$ by 
$\omega_0 = (3\alpeff/{\pi})\ 4\ \ln 2$.
For the $t$ dependence of the cross section,
an approximate power-law behaviour is predicted of the form
${\rm d}\sigma/{\rm d}\tmod \sim \tmod^{-n}$, where
$n$ depends on the parton density
functions of the proton and on the value of $\alpeff$.
In order to describe the data, the $t$ dependence of the
diffractive photon scattering simulation
was weighted by a factor $|t|^{0.73}$.
This weighted {\sc HERWIG} prediction is used only to correct the data for
resolution and acceptance effects.

Possible sources of background were estimated using MC simulations
and two of these sources were identified as non-negligible.
The background from inclusive diffractive photoproduction events
($e p \to e X Y$, where the two hadronic final states are separated by a
rapidity gap) is simulated using the {\sc PHOJET} MC event generator. 
This background contributes when a single
electromagnetic particle fakes the photon candidate and
the process is estimated to amount to $3\%$ of the measured cross section.
The background from electron pair production ($ep \rightarrow ee^+e^-X$) is
modelled using the {\sc GRAPE} event generator.
This process contributes to the selection if 
two of the leptons fake signals from the photon candidate and the scattered electron,
whereas the remaining lepton
escapes detection. This background contributes $4\%$ of the measured cross
section.

\section{Cross section measurement}

The $e p \to e \gamma Y$ differential cross sections are defined using
the formula:
\begin{equation}
 \nonumber
 \frac{{\rm d^2}\sigma_{ep \to e \gamma Y}}{{\rm d} W \ {\rm d}t}
  = \frac{N_{data} - N_{bgr}}{{\cal L} A\, \Delta W \, \Delta t},
\end{equation}
where $N_{data}$ is the number of observed events corrected for trigger
efficiency, $N_{bgr}$ the
expected contribution from background events, ${\cal L}$ the
integrated luminosity, $A$ the signal acceptance and $\Delta W$ and
$\Delta t$ the bin widths in $W$ and $t$, respectively.
The $\gamma p \to \gamma Y$ differential cross
section is then extracted from the $ep$ cross section using:
\begin{equation}
\nonumber
 \frac{{\rm d^2}\sigma_{ep \to e \gamma Y}}{{\rm d} W \ {\rm d}t}
  = \Gamma(W) \ \frac{{\rm d}\sigma_{\gamma p \to \gamma Y}}
    {{\rm d}t}(W),
\end{equation}
where the photon flux, $\Gamma(W)$, is integrated over the range
$Q^2<0.01$ GeV$^2$ according to the modified Weizs\"acker-Williams
approximation \cite{Frixione:1993yw}.
The $\gamma p$ cross section is obtained by modelling
$\sigma_{\gamma p \to \gamma Y}$ as a power-law in $W$,
whose parameters are iteratively adjusted to reproduce the measured
$W$ dependence of the $ep$ cross section.
The differential $\gamma p$ cross section in $\tmod$ is then extracted
from the $ep$ cross section by correcting for the effect of the photon flux
over the visible $W$ range.
The $\gamma p$ cross section as a function of $W$ is obtained by
first integrating the above equation over the $\tmod$ range,
and then correcting for the effect of the photon flux in each bin in $W$.

\begin{figure}
\begin{center}
\includegraphics[width=0.45\columnwidth]{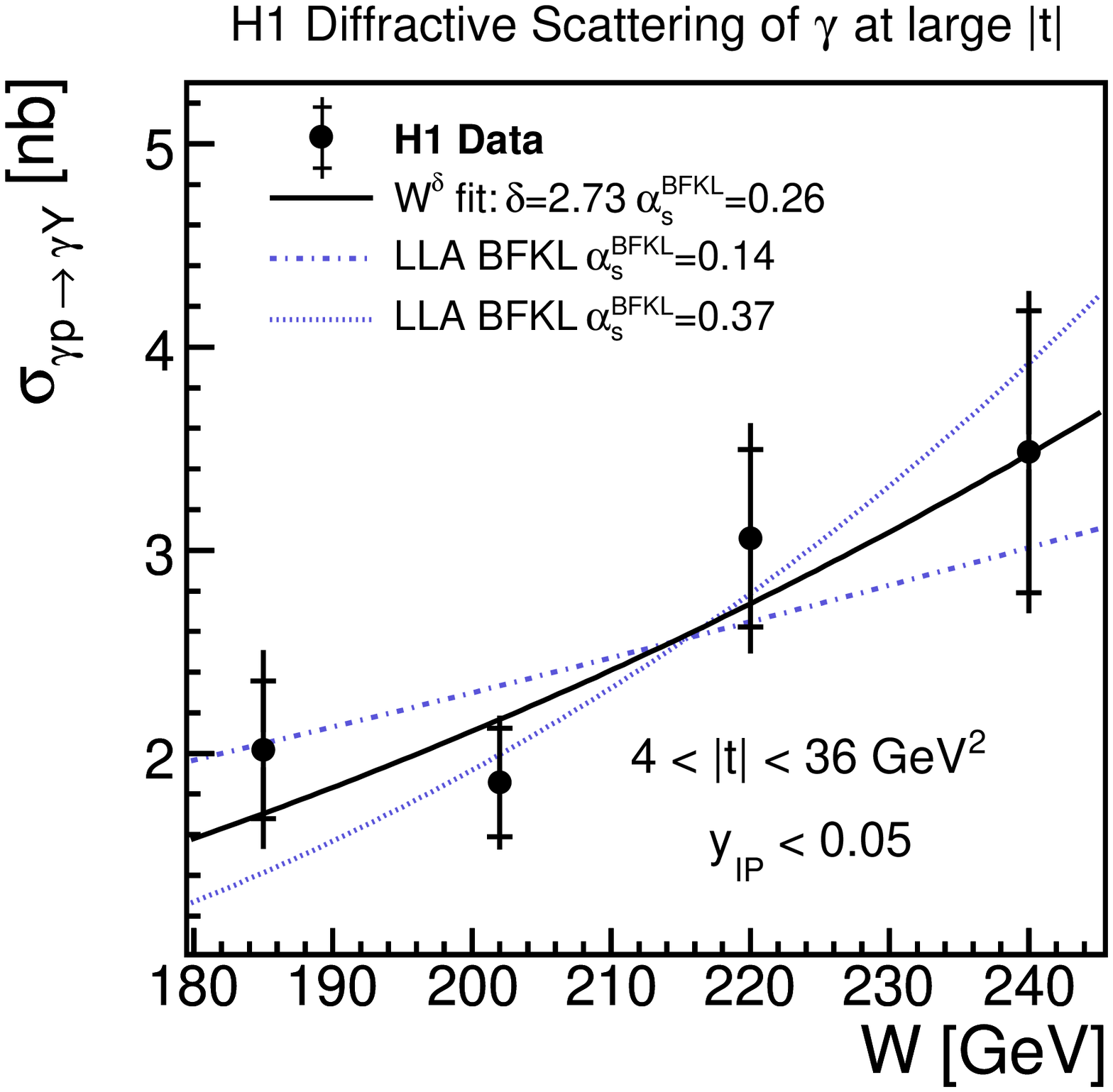}
\includegraphics[width=0.45\columnwidth]{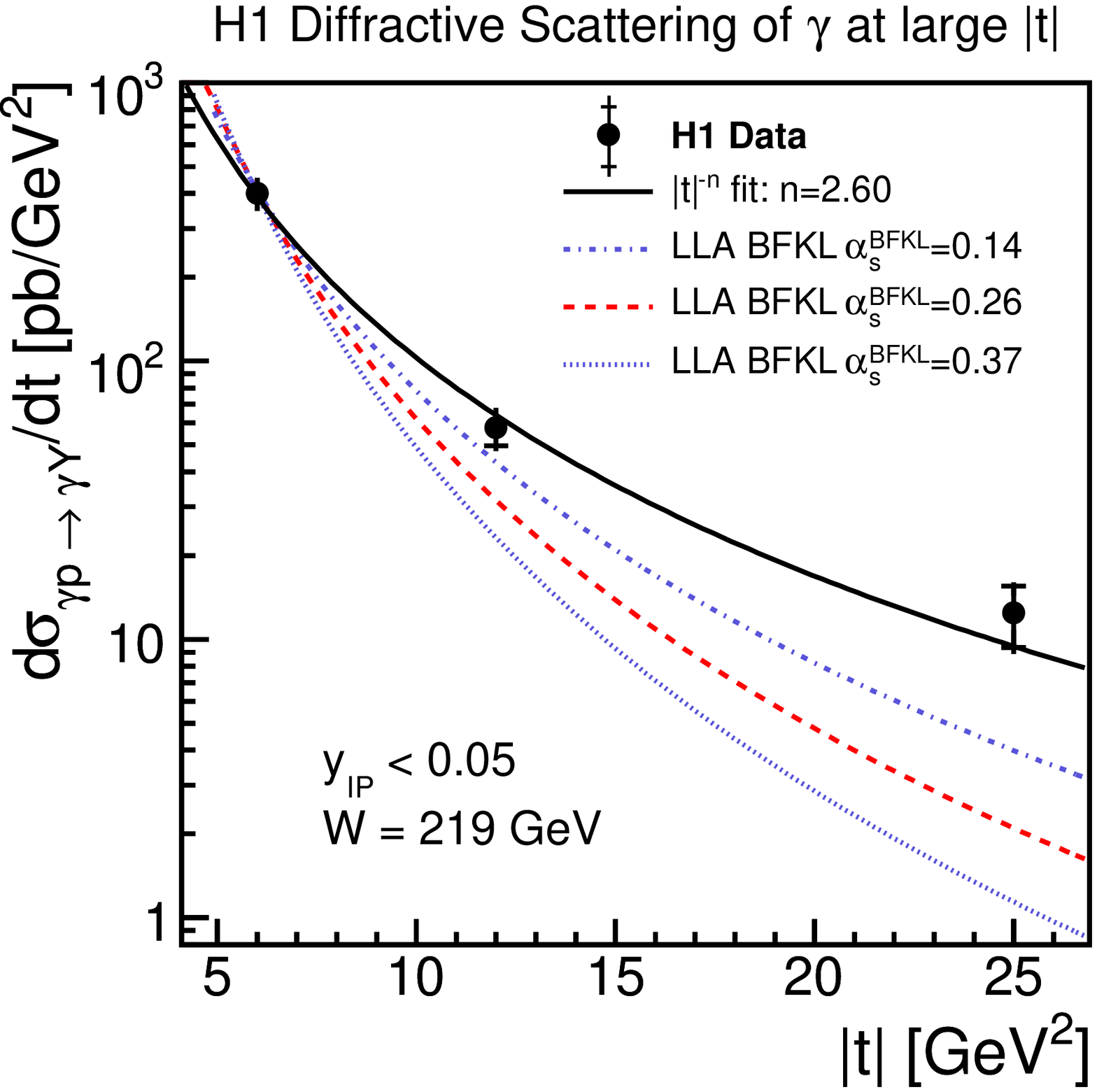}\\
{
\vspace{-1.5cm}
a)$\quad\qquad\qquad\qquad\qquad\qquad\qquad\qquad\qquad$ b)$\qquad\qquad\qquad\qquad\qquad\qquad\qquad\qquad$
}
\vspace{+1.2cm}
\caption{The $\gamma p$ cross section of diffractive scattering of
  photons as a function of
  a) $W$,
  b) $|t|$.
  The solid line represents a fit to the cross section.
  Additional curves show the LLA BFKL predictions 
  corresponding to different $\alpeff$.}
\end{center}
\label{Fig:Sigma}
\end{figure}

The systematic error on the measurement stems from experimental
uncertainties and from model dependences.
They are calculated using the weighted {\sc HERWIG}
simulation of the signal process.
The uncertainty on the PHOJET MC normalisation and the size of the
variation of the model dependence on \xpom, $\tmod$
and $M_Y$, are estimated from the measured distributions
in the data: the variation corresponds to the a range where the model still
describes the data.
Each source of systematic error is varied in the weighted HERWIG Monte
Carlo within its uncertainty. 
The largest systematic error on the cross sections comes
from the uncertainty on the $\xpom\ $ and $M_Y$ dependences.
The total systematic error on the $W$ and $\tmod$ dependence of the cross section varies
from $10\%$ to $17\%$ and from $8\%$ to $14\%$, respectively. An additional
global uncertainty of 4\% arises from the $\gamma p$ cross section
extraction procedure.
The total systematic errors are comparable to or smaller than the
statistical errors.

\section{Results and conclusions}

The cross sections are measured in the domain $175 < W < 247~\mathrm{GeV}$,
$4 < \tmod < 36$ GeV$^2$, $\ypom < 0.05$ and $\qsq<0.01\gevsq$.
The $\gamma p \rightarrow \gamma Y$ cross section as a function
of $W$ is shown in 
Fig.~3a.
A power-law dependence of the form
$\sigma \sim W^{\delta}$ is fitted to the measured cross section,
the fit yields $\wfitres$ with $\chi^2/\textrm{n.d.f.}=2.7/2$. 
The steep rise of the cross
section with $W$ is usually interpreted as an indication of the
presence of a hard sub-process in the diffractive interaction and of
the applicability of perturbative QCD.
The present $\delta$ value, measured at an average $\tmod$
value of $6.1$~GeV$^2$,
is compatible with that measured by H1 in diffractive
$J/\psi$ photoproduction of $\delta=1.29 \pm0.23 (\mathrm{stat.})\pm
0.16(\mathrm{syst.})$ \cite{Aktas:2003zi} at an average $\tmod$ of
$6.9$~GeV$^2$.
The Pomeron intercepts associated with these $\delta$ values
correspond to the strongest energy dependences
measured in diffractive processes.

The $\gamma p$ cross section differential in $\tmod$, at $W=219$ GeV,
is shown in 
Fig.~3b,
together with a fit of the form ${\rm d}\sigma/{\rm d}t \sim \left|t\right|^{-n}$,
where the fitted $\tfitres$ with $\chi^2/\textrm{n.d.f.}=1.6/1$.
The $\tmod$ dependence is harder than that
measured by H1 in the diffractive photoproduction of $J/\psi$ mesons at large $|t|$
\cite{Aktas:2003zi}.
The measured cross
sections are compared to predictions of the LLA BFKL model, using the
{\sc HERWIG} Monte Carlo with no $\tmod$ weighting.
The predictions are normalised to the integrated measured cross
section, as the normalisation is not predicted by the LLA BFKL
calculation.
The $W$ dependence of the cross section is well described by the LLA
BFKL prediction.
Using $\delta= 4\, \omega_0 = 4\ (3\alpha_S^{Fit}/{\pi})\ 4\ \ln 2$,
the measured $W$ dependence leads to $\alphasfitres$.
Predictions are shown in 
Figure 3 
also for
the values $\alpeff=0.14$ and $0.37$, corresponding to one standard deviation
of $\alpha_S^{Fit}$. 

As shown in 
Figure 3b, 
the LLA BFKL calculation
for $\alpeff=0.14, ~0.26$ and $0.37$, all of which give
a reasonable description of the $W$ dependence, predict steeper $\tmod$
distributions than is measured in the data.
The same effect cannot be established for the exclusive $\rho$
measurement \cite{Aktas:2006qs}, where the measured $t$ range is limited to
$\tmod < 8$ GeV$^2$, although an underestimate of the cross section was
observed at the largest values of $\tmod$.
The present situation is in contrast with the analysis of $J/\psi$
production \cite{Aktas:2003zi,Chekanov:2002rm}, 
where the $\tmod$ dependence
was found to be well described by the LLA BFKL prediction over a similar
range in $t$.


\begin{footnotesize}



%

\end{footnotesize}


\end{document}